\documentclass[useAMS]{mn2e}
\usepackage{times}
\usepackage{psfig}

\newtheorem{rule-def}[theorem]{Rule}

\newcommand{\be}{\begin{equation}}
\newcommand{\ee}{\end{equation}}
\newcommand{\bea}{\begin{eqnarray}}
\newcommand{\eea}{\end{eqnarray}}

\begin{document}

\title[Tolman-Bayin type static charged...]{\bf Tolman-Bayin type static
charged fluid spheres in general relativity}

\author[Ray \& Das]{Saibal Ray$^{1,2}$ \& Basanti Das$^{3}$\\
$^1$Department of Physics, Barasat Government College, Barasat 700 124, North 24
Parganas, West Bengal, India \\
$^2$Inter-University Centre for Astronomy and Astrophysics, PO Box 4, Pune 411 007,
India; e-mail: saibal@iucaa.ernet.in\\
 $^3$Belda Prabhati Balika Vidyapith, Belda, Midnapur 721 424, West Bengal, India.
}

\maketitle

\begin{abstract}
{In a static spherically symmetric Einstein-Maxwell spacetime the
class of astrophysical solution  found out by Ray and Das (2002)
and Pant and Sah (1979) are revisited here in connection to the
phenomenological relationship between the gravitational and
electromagnetic fields. It is qualitatively  shown that the
charged relativistic stars of Tolman (1939) and Bayin (1978) type
are of purely electromagnetic origin. The existence of this type
of astrophysical solutions is a probable extension of Lorentz's
conjecture that electron-like extended charged particle possesses
only `electromagnetic mass' and no  `material mass'.}

\end{abstract}


\section{Introduction}

The study of the interior of stars is always fascinating to the
astrophysicists, specially in connection to general theory of
relativity. This is obvious because of the fact that towards the
late stages of stellar evolution, general relativistic effects
become much important. One of the remarkable works in this
direction was that of the Tolman (1939) solutions. Tolman
extensively studied the stellar interior and provided a class of
explicit solution in terms of known analytic functions for the
static, spherically symmetric equilibrium fluid distribution.
Subsequently Wyman (1949), Leibovitz (1969) and Whitman (1977)
generalized some of Tolman's solutions. Bayin (1978) also
obtained some more new analytic solutions related to static fluid
spheres using the method of quadratures.

Recently we (Ray \& Das, 2002) have obtained the charged
generalization of Bayin's work (1978) motivated by the idea that
in stellar astrophysics the coupled Einstein-Maxwell field
equations may have some physical implications. In connection to
singularity problem it is observed that in the presence of
charge, the gravitational collapse of a spherically symmetric
distribution of matter to a point singularity may be avoided. The
mechanism is such that the gravitational attraction is
counterbalanced by the repulsive Coulombian force in addition to
the thermal pressure gradient due to fluid. Also, it is seen that
the presence of the charge function serves as a safety valve,
which absorbs much of the fine-tuning, necessary in the uncharged
case (Ivanov, 2002). Thus, the problem of coupled charge-matter
distributions in general relativity has received considerable
attention.

The present paper is based on the simple investigation of the
solutions already obtained by us (Ray \& Das, 2002) and Pant \&
Sah (1979) in connection to the electromagnetic origin of the
gravitational mass. It is worthwhile to mention here that there
is a fairly long history of investigations about the nature of
the  mass of electron. Einstein (1919) himself believed that
``... of the energy constituting matter three-quarters is to be
ascribed to the electromagnetic field, and one-quarter to the
gravitational field'' whereas Lorentz's (1904) conjecture of
extended electron was that ``there is no other, no `true' or
`material' mass,'' and thus provides only 'electromagnetic masses
of the electron'. Wheeler (1962) also believed that electron has
a `mass without mass'. Feynman (1964) termed this type of models
as `electromagnetic mass models' in his classic volume. Starting
from 60's in the last century several authors (e.g., Florides,
1962; Cooperstock \& De La Cruz, 1978; Tiwari et al., 1984;
Gautreau, 1985; Gr{\o}n, 1986; Ponce de Leon, 1987; and the
references therein) took up the problem again and studied
electromagnetic mass models for the static spherically symmetric
charged perfect fluid distribution in the framework of general
relativity. Very recenty the idea is extended to the
Einstein-Cartan theory and Kaluza-Klein theory by adding torsion
and higher dimension respectively (Tiwari \& Ray, 1997; Ponce de
Leon, 2003). Most of these workers exploit an equation of state
$\rho + p = 0$ where, in general, the matter density $\rho>0$ and
pressure $p<0$. This type of equation of state implies that the
matter distribution under consideration is in tension and hence
the matter is known in the literature as a `false vacuum' or
`degenerate vacuum' or `$\rho$-vacuum' (Davies, 1984; Blome \&
Priester, 1984; Hogan, 1984; Kaiser \& Stebbins, 1984).

It is interesting to note that in the present study, even though
the solutions related to pressure and density in general follow
the ordinary equation of state, viz., $\rho + p \neq 0$ but
ultimately in connection to electromagnetic mass models it turns
out to be the exotic kind of equation of state (Davies, 1984;
Blome \& Priester, 1984; Hogan, 1984; Kaiser \& Stebbins, 1984)
in both the cases of Bayin and Tolman solutions. We have
investigated here that related to this type of vacuum- or
imperfect-fluid equation of state the charged analogue of Bayin
(1978) and Tolman (1939) type astrophysical class of solution
show the electromagnetic field dependency of gravitational mass.
Therefore, the existence of this type of solutions, in our
opinion, is a probable extension of Lorentz's conjecture in
connection to astrophysical models.

\section{Einstein-Maxwell field equations }
We write the line element for static spherically symmetric
spacetimes in the form
\begin{eqnarray}
ds^{2} = A^{2} dt^{2} - B^{2} dr^{2} - r^{2} (d\theta^{2} +
sin^{2}\theta d\phi^{2}).
\end{eqnarray}
in the standard coordinates $x^i=(t,r,\theta,\phi)$, where the
quantities $A(r)$ and $B(r)$ are the metric potentials.

\noindent
The Einstein-Maxwell field equations, for the metric (1) in the
comoving coordinates read as
\begin{eqnarray}
\frac{1}{B^2}\left(\frac{2B'}{Br} - \frac{1}{r^2}\right) + \frac{1}{r^2}
 = 8\pi \rho + \frac{q^2(r)}{r^4},\label{field1}\\
\frac{1}{B^2}\left(\frac{2A'}{Ar} + \frac{1}{r^2}\right) - \frac{1}{r^2}
 = 8\pi p - \frac{q^2(r)}{r^4},\label{field2}\\
\frac{1}{B^2}\left[\frac{A''}{A} - \frac{A'B'}{AB} + \frac{1}{r
}
\left(\frac{A'}{A} - \frac{B'}{B}\right)\right] = 8\pi p + \frac{q^2(r)}{r^4},
\label{field3}
\end{eqnarray}
where the prime denotes differentiation with respect to radial
coordinate $r$ only. In the equation (\ref{field1}) -
(\ref{field3}), the quantities $\rho$, $p$ and $q$ represent the
energy density, isotropic pressure and total electric charge
respectively. The total charge within a sphere of radius $r$ is
defined as
\begin{eqnarray}
q(r) = 4\pi \int_0^r J^0 r^2 AB dr,
\end{eqnarray}
$J^i$ being the $4$-current takes here the form, via the
electromagnetic field $F^{01}$, as
\begin{eqnarray}
F^{01} = \frac{q(r)}{AB{r^2}} .
\end{eqnarray}
Now, eliminating $p$ from equations (3) and (4) and assuming
$A'/Ar = C(r)$ one can get
\begin{eqnarray} \nonumber
\left(\frac{1}{B^3r^2} + \frac{C}{B^3}\right)dB - \left(\frac{1}{B^2}\right)dC&&\\
~~~~~~~~~~~~ - \left(\frac{B^2 - 1}{B^2r^3} + \frac{C^2r}{B^2} - \frac{2q^2}{r^5}\right)dr=0.
\label{nop}
\end{eqnarray}
which is a Pfaffian differential equation in three dimensions
having the general form as
\begin{eqnarray}
f_{1}(B, C, r) dB + f_{2}(B, C, r) dC + f_{3}(B, C, r) dr = 0.
\end{eqnarray}

\section{Electromagnetic mass models for static charged fluid spheres}
\subsection{Bayin's class of solution}\label{class}

The Pfaffian differential equation (\ref{nop}) can be solved in
different ways as shown by us (Ray \& Das, 2002) in details. It
is shown that in terms of $B(r)$ when $C(r)$ is known, the
Pfaffian differential equation (\ref{nop}) becomes
\begin{eqnarray}
\frac{dB}{dr} = \left[\frac{1 - \frac{2q^2}{r^2}}{\left(C +
\frac{1}{r^2}\right)r^3}\right]B^3 + \left[\frac{C^{2}r - \frac{1}{r^3} +
\frac{dC}{dr}}{C + \frac{1}{r^2}}\right]B.
\label{dbdr}
\end{eqnarray}
Also, in terms of $C(r)$ when $B(r)$ is given, the Pfaffian
differential equation (\ref{nop}) modifies to
\begin{eqnarray}
\nonumber
\frac{dC}{dr} = \left(\frac{1}{Br^2} \frac{dB}{dr} - \frac{B^2 - 1}{r^3}\right)
&+& \left(\frac{1}{B} \frac{dB}{dr}\right) C\\
- {C^2}r &+& \frac{2B^2q^2}{r^5},
\label{dcdr}
\end{eqnarray}
which is a Riccati equation for $C(r)$ with known value of charge $q$.

By solving these differential equations (\ref{dbdr}) and
(\ref{dcdr}), and also some other simple cases we (Ray \& Das,
2002) obtained the solutions for Einstein-Maxwell field equations
related to Bayin (1978) type astrophysical class of models. The
solutions thus obtained for the parameters $A$, $B$, $\rho$, $p$
and $q$ respectively the gravitational potentials, energy
density, isotropic pressure and electric charge are involved with
several integration constants. Some of these may, in principle,
be determined by matching of the interior solution to the
exterior Reissner-Nordstr\"{o}m metric at the boundary $r=a$ of
the spherical matter distribution. The exterior
Reissner-Nordstr\"{o}m metric is given by
\begin{eqnarray}
\nonumber
ds^2 = \left(1 - \frac{2m}{r} + \frac{q^2}{r^2}\right)dt^2
&-& \left(1 -  \frac{2m}{r} + \frac{q^2}{r^2}\right)^{-1}dr^2 \\
&-& r^2(d\theta^2 + sin^2\theta d\phi^2).
\label{ds2rn}
\end{eqnarray}
Now, considering the metric components $g_{00}$, $g_{11}$ and
$\frac{\partial{g_{00}}}{\partial{r}}$ to be continuous across
the boundary $r = a$ of the sphere and assuming for the total
charge on the sphere
\begin{eqnarray}
q(a) = Ka^n,
\label{poly}
\end{eqnarray}
one can get the following cases of the gravitational mass (vide
equations (53), (56), (61), (65), (69) and (72) of Section $4$ in
Ray \& Das (2002)) in the explicit forms with electric charge.

\noindent Case I (i): For $n=1$
\begin{eqnarray}
m = q^2 + a_0a_1\left(\frac{q}{K}\right)^2 + {a_1}^2\left(\frac{q}{K}\right)^3,
\end{eqnarray}
(ii): For $n = 3$
\begin{eqnarray}
m = {a_1}^2\left(\frac{q}{K}\right) + 4K^2\left(\frac{q}{K}\right)^{5/3} +  a_0a_1\left(\frac{q}{K}\right)^{2/3},
\end{eqnarray}
Case II (i): For $n = 1$
\begin{eqnarray}
\nonumber
m =\frac{1}{{W_0}^2}\left[{W_0}^2(1 - 2 K^2) +
\left(\frac{q}{K}\right)^2\left(C_1 - \frac{{q}^2}{{K}^2}\right)\right]^{1/2}\\
\times \left[\frac{q}{K}\right]^3,
\end{eqnarray}
(ii): For $n = 3$
\begin{eqnarray}
\nonumber
m &=& \frac{1}{{W_0}^2}\left[{W_0}^2 + C_1 \left(\frac{q}{K}\right)^{2/3} +
({{W_0}^2}K^2 - 1) \left(\frac{q}{K}\right)^{4/3}\right]^{1/2}\\
& &\times \left[\frac{q}{K}\right]- 2K^{1/3}q^{5/3},
\end{eqnarray}
Case III: For $n = 1$
\begin{eqnarray} m = \frac{1 + K^2}{C_3} = (K^2
- 1)\left[C_3 \left(\frac{q}{K}\right) - 2\right]
\left[\frac{q}{K}\right],
\end{eqnarray}
Case IV: For $n = 1$
\begin{eqnarray} m = 3C_6^2\left[3C_5 +
\left(\frac{q}{K}\right)^3\right] \left[\frac{q}{K}\right]^4,
\end{eqnarray}
where $K$, $a_0$, $a_1$, $W_0$, $C_1$, $C_3$, $C_5$ and
$C_6$ all are constant quantities. Now, $m$ for the cases I(i),
I(ii) and IV are as usual positive whereas for the rest of the
cases II(i), II(ii) and III the conditions for positivity are
$(1/2)> K^2 >(q^2/C_1)$, $ K>(1/W_0)$ and $1<K<(C_3 q/2)$
respectively.

It is observed from the explicit forms of the above set of
expressions that the effective gravitational mass $m$, along with
the central pressures and densities at $r=0$ (vide equations
(22), (27), (31), (40), (45) and (21), (28), (32), (39)
respectively in Ray \& Das (2002)), is related to the charge $q$
of equation (\ref{poly}) of the spherical system. Therefore,
vanishing of the charge makes all the physical quantities
including the gravitational mass also to vanish. This means that
the gravitational mass originates from the electromagnetic field
alone. Thus, the gravitational mass is purely `electromagnetic
mass' (Lorentz, 1904) and this type of model is known as
`electromagnetic mass model' in the literature (Feynman, 1964).
It is relevant to note here that this particular important
feature of the solution set, viz., the electromagnetic nature of
the gravitational mass is obviously not available in the
uncharged case of Bayin (1978) and thus indicates that the
presence of charge allows for a wider range of behaviour.

\subsection{Tolman's solution VI}
In the introduction we have mentioned that motivated by the work
of Tolman (1939), a similar kind of new class of solution was
found by Bayin (1978) and hence in view of the results of
sub-section \ref{class} related to Bayin's work it will be
interesting to examine the solutions of Tolman whether they are
also a member of electromagnetic mass models. As a ready-made
example we would like to present here the solution obtained by
Pant \& Sah (1979) to meet our, at least partial, requirement.
For a static spherically symmetric distribution of charged fluid
the solution set (vide equations (10a), (10b), (10c) and (11) in
Pant \& Sah (1979)) is as follows.
\begin{eqnarray}
A^2 = e^{\nu} = b r^{2n},\label{asquare}
\end{eqnarray}
\begin{eqnarray}
B^{-2} = e^{-\lambda} = c,\label{bsquare}
\end{eqnarray}
\begin{eqnarray}
\rho = \frac{1}{16\pi r^2}[1 - c(n - 1 )^2],\label{rhoc}
\end{eqnarray}
\begin{eqnarray}
p = \frac{1}{16\pi r^2}[c(n+1)^2 -1],\label{pc}
\end{eqnarray}
\begin{eqnarray}
\sigma = \pm \frac{1}{4\pi r^2}[\frac{c}{2}\{1 - c (1 + 2n -
n^2)\}]^{1/2},
\end{eqnarray}
\begin{eqnarray} E^2 =  \frac{1}{2 r^2}[1 -
c(1 + 2n - n^2)],
\end{eqnarray} where
\begin{eqnarray} b =
{a}^{-2n}\left[1 - \frac{2m}{a} + \frac{{q}^2}{{a}^2}\right],
\end{eqnarray}
\begin{eqnarray} c = \left[1 - \frac{2m}{a} +
\frac{{q}^2}{{a}^2}\right] = \left[1 -
\frac{{2q}^2}{{a}^2}\right]( 1 + 2n - n^2)^{-1}.
\end{eqnarray}
The above set of solutions, in view of $c$,
with $\Lambda=0$ and $B=0$ represents the charged analogue of
Tolman's (1939) solution VI and thus in the absence of the total
charge $q$ reduces to the neutral one (the sub-case $C$ of
uncharged fluid sphere in the Pant \& Sah (1979)). Now, the
equation (\ref{field1}) can be expressed in the form
\begin{eqnarray}
B^{-2} = 1 - \frac{2m(r)}{r},\label{bminus2}
\end{eqnarray}
where the gravitational mass $m(r) = M(r) + \mu(r)$ being defined
as
\begin{eqnarray}
M(r) = 4 \pi \int_{0}^{r} \rho r^2 dr
\end{eqnarray} and
\begin{eqnarray} \mu(r) = \int_{0}^{r} ({q^2}/{2 r^2}) dr,
\end{eqnarray}
respectively the Schwarzschild mass and the mass equivalence of
electromagnetic field. Hence, the total gravitational mass,
$m(r=a)$, can be calculated as
\begin{eqnarray}
m = \frac{na^2(2 - n) + 2q^2}{2(1+2n-n^2)a}.
\end{eqnarray}
If we now make the specific choice $n=0$ for the parameter $n$
appearing in the above solution set then one get the following
expressions.
\begin{eqnarray} A^2 =
\left[1 - \frac{2m}{a} + \frac{{q}^2}{{a}^2}\right],\label{a2}
\end{eqnarray}
\begin{eqnarray} B^{-2} = \left[1 - \frac{2m}{a} +
\frac{{q}^2}{{a}^2}\right] = \left[1 -
\frac{2{q}^2}{{a}^2}\right],
\label{b2}
\end{eqnarray}
\begin{eqnarray}
\rho(r) = \frac{1}{8\pi r^2}\left[\frac{q^2}{a^2}\right],\label{rhor}
\end{eqnarray}
\begin{eqnarray}
p(r) = - \frac{1}{8\pi r^2}\left[\frac{q^2}{a^2}\right],\label{pr}
\end{eqnarray}
\begin{eqnarray}
\sigma(r) = \pm \frac{1}{4\pi r^2}\left[\frac{q}{a}\right]\left[1
- \frac{{2q}^2}{{a}^2}\right]^{1/2},
\end{eqnarray}
\begin{eqnarray}
E(r) = \frac{q}{a r},
\end{eqnarray}
and
\begin{eqnarray}
m = \frac{q^2}{a}.
\end{eqnarray}
Thus, for vanishing electric charge all the physical quantities
including gravitational mass vanish and the spacetime becomes
flat. It is interesting to note that, in the present situation,
the equations (\ref{rhor}) and (\ref{pr}) related to the
isotropic pressure and matter density provide an equation of
state $\rho + p =0$, which is known as the vacuum- or
imperfect-fluid equation of state. As is evident, from the
equations (\ref{rhoc}) and (\ref{pc}), this is not true for the
general case when $n\neq 0$ and can be read as
\begin{eqnarray}
\rho + p = \frac{1}{4\pi r^2}\left[\frac{n(a^2 - 2q^2)}{a^2(1 +
2n - n^2)}\right].
\end{eqnarray}
Hence starting from a perfect fluid type equation of state via $n
= 0$ we are arriving at the imperfect-fluid type equation of
state and thus $n$ here is taking a definite and peculiar role
for deciding the form of the equation of state. This particular
aspect is also true via the equations (\ref{field1}) and
(\ref{field2}) for the equations (\ref{asquare}) and
(\ref{bsquare}) which reduce to the equations (\ref{a2}) and
(\ref{b2}) respectively, with $n = 0$ when we get $\nu + \lambda
= 0$. This again, for the Reissner-Nordstr\"{o}m metric (equation
(\ref{ds2rn})) related to the spherically symmetric static
charged fluid distribution, can be expressed in the form
$g_{00}g_{11} = - 1$. Thus, in view of equations (\ref{field1})
and (\ref{field2}), we see that for the boundary condition $\nu +
\lambda = 0$ one can get  $\rho + p = 0$ and vice versa, so that
$\lambda = -\nu \leftrightarrow p = - \rho$.\footnote{A
coordinate-independent statement of the relation $g_{00}g_{11} =
- 1$ and hence $\nu + \lambda = 0$ is given by Tiwari et al.
(1984).} This result means that if $\rho>0$ then must be $p<0$
for the inside of the fluid sphere though, in general, $\rho$ and
$p$ are positive for the condition $0 \leq n\leq 1$. This
partially admits the comment by Ivanov that ``... electromagnetic
mass models all seem to have negative pressure''. Partially
because, in our opinion, there are some examples of
electromagnetic mass models where positive pressures are also
available will be shown elsewhere.

\section{Conclusions}
We have revisited in the present paper the work already done by
us (Ray \& Das, 2002) and that one of Pant \& Sah (1979)
motivated by the fact that the gravitational mass $m$ for a
charged matter distributions always can be seen to be divided
into two parts, viz., (i) the Schwarzschild mass $M(r)$ and (ii)
the mass equivalence of electromagnetic field $\mu(r)$ as is
evident from the equation (\ref{bminus2}). Thus, the total mass
is increasing due to electromagnetic energy (Florides, 1964;
Mehra, 1980) which is obviously an extra feature in comparison to
the neutral case. Keeping this aspect in mind we wanted to
examine whether the gravitational mass obtained by us (Ray \&
Das, 2002) in one of our previous papers is of purely
electromagnetic origin or not. We have, in the present simple
investigation, shown that the charged generalized solutions of
Bayin (1978) is purely of electromagnetic origin. In this
connection, we have also shown, by citing the solution of Pant \&
Sah (1979), that the charged generalization of Tolman's solution
VI (1939) yields electromagnetic mass models which, of course,
needs further investigations with a direct study of the Tolman's
whole set of solutions by inclusion of charge.

We would also like to mention here that the works done by
different investigators on electromagnetic mass models so far are
mainly concerned with the structure of the classical electron
(special references are Gautreau, 1985 and Tiwari et al, 1986).
Even though Tiwari et al. (1986) find astrophysically interesting
Lane-Emden equations in connection to electromagnetic mass models
but at the same time, instead of studying the stellar structures,
they apply the radii of some of the models for the comparison
with the classical electron radius. This particular aspect of
electromagnetic mass models related to Lane-Emden equations in
the astrophysical context needs further investigations.

As is mentioned in Ray \& Das (2002), to justify the present work
with a charged fluid distribution, that even though the
astrophysical systems are by and large electrically neutral,
recent studies do not rule out the possibility of the existence
of massive astrophysical systems that are not electrically neutral
(Treves \& Turella, 1999). The mechanism, though not completely
understood, is mainly related to the acquiring a net charge by
accretion from the surrounding medium. On the other hand, there
are some other views of acquiring charge by a compact star during
its collapse from the supernova stage. In this regard it will be
worth mentioning that to study the effect of electric charge in
compact stars Ray et al. (2003) assume an {\it ansatz} such that
$\sigma = \alpha \rho$ where $\alpha$ is related to the charge
fraction $f$ as $\alpha = 869.24 f$ and show by numerical
calculation that in order to see any appreciable effect on the
phenomenology of the compact stars, the total electric charge is
to be $\sim 10^{20}$ Coulomb. Therefore, in our opinion, even
such a remote possibility gives enough scope to theoretical
speculations and hence the corresponding modeling and
investigations become as much pertinent for these cases as for
the established neutral systems.

\section*{Acknowledgments}
One of the authors (SR) is thankful to the authority of
Inter-University Centre for Astronomy and Astrophysics, Pune,
India, for providing Associateship programme under which a part
of this work was carried out.




\end{document}